\documentclass[12pt]{article}
\usepackage{graphicx}

\begin{document}

\begin{titlepage}
\centerline{\large \bf MAXIMUM ENTROPY METHOD AND OSCILLATIONS}
\centerline{\large \bf IN THE DIFFRACTION CONE}

\vskip 1.0cm

\centerline{O. Dumbrajs$^{1\dagger}$, J. Kontros$^{2\ddagger}$,
A. Lengyel$^{2\circ}$}

\vskip .5cm

\centerline{$^{1}$ \sl Department of Engineering Physics and Mathematics,}
\centerline{\sl Helsinki University of Technology, Rakentajanaukio 2,}
\centerline{\sl FIN-02150 Espoo, Finland}

\vskip .3cm

\centerline{$^{2}$ \sl Institute of Electron Physics, Universitetska 21,}
\centerline{\sl 88000 Uzhgorod, Ukraine}

\vskip 1cm

\begin{abstract}
The maximum entropy method has been applied to investigate the
oscillating structure in the $\overline{p}p-$ and $pp-$elastic
scattering differential cross-section at high energy and small
momentum transfer. Oscillations satisfying quite realistic
reliability criteria have been found.
\end{abstract}

\vskip 10cm

\hrule

\vskip .9cm

\noindent

$ \begin{array}{ll}
^{\dagger}\mbox{{\it e-mail address:}} &
   \mbox{dumbrajs@csc.fi}
\end{array}
$

$ \begin{array}{ll}
^{\ddagger}\mbox{{\it e-mail address:}} &
 \mbox{jeno@kontr.uzhgorod.ua}
\end{array}
$

$ \begin{array}{ll}
^{\circ}\mbox{{\it e-mail address:}} &
 \mbox{sasha@len.uzhgorod.ua}
\end{array}
$

\vfill
\end{titlepage}
\baselineskip=14pt

\section{Introduction}

The presence of anomalous structure in the differential cross
sections was always indicative of some important physical feature
typical of the particles and their interactions. Such are the
dip-bump structure and "fine structure" in the differential cross
sections related to small and large values of the impact
parameter. The first phenomenon is connected with absorption at
small impact parameters, while the second one to the mesonic (or
quark-antiquark) "cloud" (or "tail") of the colliding particles.
In addition to this, localized fluctuations have been observed
\cite{WH73,BSW93} in the differential cross-section at fixed
$\sqrt{s}$. The oscillating structure of the diffraction peak in
the differential $\overline{p}p-$ and $pp-$elastic cross sections
has been observed for the first time in \cite{WH73} in the ISR
data \cite{BB72} and later \cite{BSW93} in UA4/2 \cite{CA93}
experiment by normalizing the differential cross-section to the
smoothly varying background in the impact parameter
representation. In \cite {AB76} an attempt was made to relate the
observed structure near $|t|=0.1$ $GeV^2$ with the variation of
the opacity in $b-$space, probably reflecting the density
oscillation in matter. The possible existence of oscillations with
even smaller periods was discussed by several authors
\cite{BH94,PG97,KFT97}. As pointed out in \cite{BSW93}, the
Coulomb interference region is essential for the clarification of
the presence of these oscillations in $\overline{p}p-$ and
$pp-$elastic scattering cross sections. The anomalous behaviour
observed in the experimental data may be a manifestation of the
presence of coherent states in the hadronic matter. In ref.
\cite{BB72} this anomalous behaviour was described by a
phenomenological formula and interpreted as an unusual phenomenon
giving rise to new physics in hadron diffraction.

The smallness of the relevant periods of oscillations indicates
the presence of the long-range forces. Recently in \cite{KFT97} it
was suggested that the long-range forces at $t\rightarrow 0$ give
a significant contribution to the scattering amplitude and may be
responsible for the structure observed in $\frac{d\sigma }{dt}$ at
small $-t\sim 10^{-3}$ $GeV^2$. It has been conjectured in
\cite{PG97} that the oscillations of the differential
cross-section observed in \cite{CA93} at $\sqrt{s}=541$ $GeV$ are
periodic in $\sqrt{t}$ with the period of $2\cdot 10^{-2}$ $GeV$
which is compatible with the Auberson-Kinoshita-Martin type of
oscillations \cite{AKM71}.

In ref. \cite{KL95} we proposed an entirely different method of identifying
oscillations in the $\overline{p}p-$ and $pp-$elastic scattering. Our method
was based on the use of overlapping bins of local slopes and we indicated
some characteristic oscillation periods. It is quite obvious that, in order
to determine the nature and periods of oscillations, one has first of all to
increase the reliability of initial information contained in experimental
data by suppressing the influence of statistical fluctuations. This problem
can be settled by means of the well known method of maximum entropy \cite
{Max91}, used in many branches of physics. For example, recently it has been
applied \cite{KD97} to the electron density profile reconstruction from
multichannel microwave interferometer data at the \textit{W7-AS} fusion
experiment.

\section{The method of maximum entropy}

Consider the mathematical essence of the maximum entropy method. The task is
formulated as follows. Provided there are $N$ measured data points $y_k$
with errors $\varepsilon _k$ and no theory available, the question arises of
how to choose the most plausible function $Y_k$ among all possible functions
describing this data set. The philosophy is to choose the solution which
contains the least amount of information in order to avoid false features.
Mathematically this means that one has to choose such a function $Y_k$ that
the functional $F$ is minimized:

\begin{equation}
F=\chi ^2+\lambda R.  \label{eq1}
\end{equation}
Here $\chi ^2$ is the least squares term which provides the consistency of
the function $Y$ with the data set $y$, $R$ is a regularization functional
which helps to choose the correct function containing the least amount of
information, and $\lambda $ is the weighting factor between the least
squares term and the regularization functional. The value of $\lambda $ has
to be chosen such that the data is neither under- nor overestimated, i.e., $%
\chi ^2$ should be equal to the number of data points $N$. The $\chi ^2$ is
calculated by means of the standard formula

\begin{equation}
\chi ^2=\sum_{k=1}^N\frac{\left( Y_k-y_k\right) ^2}{\varepsilon _k^2}.
\label{eq2}
\end{equation}
It is very important to use realistic error values with methods based on
regularization functional, in order to get realistic results. In the maximum
entropy method the regularization functional is the configurational entropy
of the distribution:

\begin{equation}
R=-\sum_{k=1}^Np_k\log \frac{p_k}{m_k},  \label{eq3}
\end{equation}
where

\begin{equation}
p_k=\frac{Y_k}{\sum_{k=1}^N{Y_k}},  \label{eq4}
\end{equation}
$m_k$ - is the $k$-th value of the discretised \textit{a priori}
distribution. The basic principle of the maximum entropy method is
that out of all probability distributions which satisfy given
constraints, i.e., fits to the data, one should choose the
distribution which is closest to the given \textit{a priori}
probability distribution $m_k$, and, if this is not specified, one
should choose the distribution that is closest to the uniform
distribution. Here should be pointed out that unknowns $Y_k$ are
assumed to be independent of each other which means that the
distribution does not have to be continuous.

\section{Application of the maximum entropy method to the differential
cross-section}

The question of the role of statistical errors in measured
differential cross-section $\frac{d\sigma }{dt}\left( s,t\right) $
for $\overline{p}p-$ and $pp-$scattering at fixed $s$ within a
broad $t$-interval is of substantial importance, because these
errors for each data point may be of the same order of magnitude
as the "fine" anomalies of $\frac{d\sigma }{dt} $ observed in
experiments \cite{BB72,CA93,Sch81}. The second problem is related
to ambiguous use of different models and theoretical predictions
\cite{WH73,BSW93,BH94} in fitting the experimental "curves". From
the standpoint of the "observed" oscillations, several methods are
now being extensively used. One of the methods is based on
theoretical description of smooth $\frac{d\sigma }{dt}$ and
subsequent extraction of oscillations (anomalies) by subtracting
the slowly varying "background" from the experimental differential
cross sections. In our new approach we first process the
experimental $\left( \frac{d\sigma }{dt}\right) ^{exper}$ values
according to the "maximum entropy criterion" to obtain

\begin{figure}
\begin{center}
\includegraphics[scale=0.5]{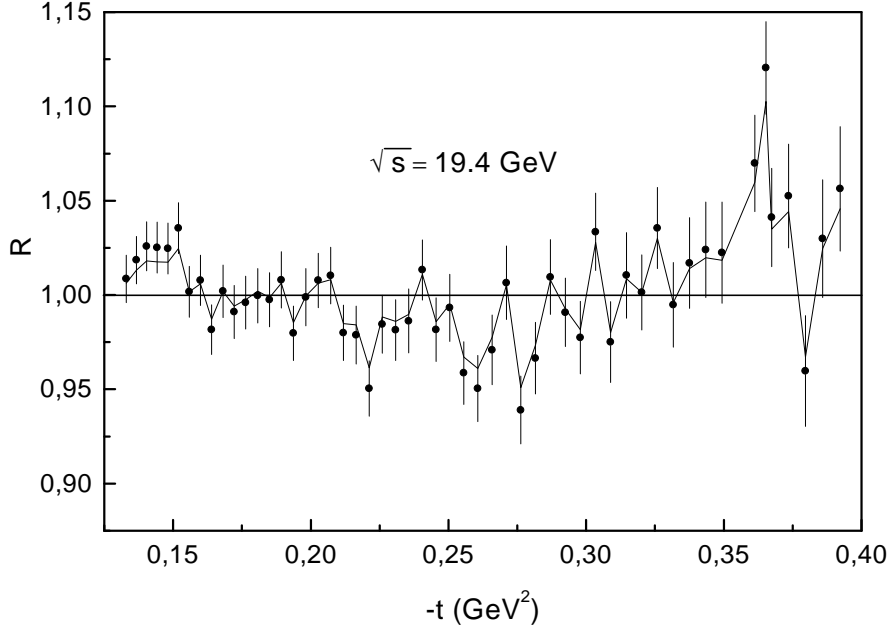}
\caption[]{Experimental values of the data at $19.4$\ $GeV$\
\cite{Sch81} calculated by means of Eq. (\ref{eq10}) (points with
error bars), maximum entropy fits with original errors (solid
curve; $\chi ^2/ndf\cong 1.08$ and $\lambda =1.5\cdot 10^6$).}
\label{fig1}
\end{center}
\end{figure}

\begin{equation}
Y_k=\left( \frac{d\sigma }{dt}\right) _k^{entr}.  \label{eq5}
\end{equation}
For this purpose we minimize the functional (\ref{eq1}), where:

\begin{equation}
y_k=\left( \frac{d\sigma }{dt}\right) _k^{exper}  \label{eq6}
\end{equation}
and $\varepsilon _k$ are experimental errors of $y_k=\left(
\frac{d\sigma }{dt}\right) _k^{exper}$. We use the parametrization
of $\frac{d\sigma }{dt}$ from \cite{Desg97} which well describes
background to calculate the $m_k$ values needed in (\ref{eq3})

\begin{equation}
m_k=\left( \frac{d\sigma }{dt}\right) _k^{theor}=\frac 1{16\pi }\left|
F_N+F_C\right| ^2,  \label{eq7}
\end{equation}
\begin{equation}
F_N=\sigma _{tot}\left( i+\rho \right) \exp \left[ bt_k+\gamma
\left( 1- \sqrt{1-t_k/t_0}\right) \right] .  \label{eq8}
\end{equation}
Here $\sigma _{tot}$, $\rho$, $b$, $t_0$ and $\gamma $ are fitted
parameters and $F_C$ is the standard Coulomb amplitude which can
be calculated from \cite{WY68}. We investigate the experimental
differential $\overline{p}p-$ and $pp-$scattering cross sections
at different energies covering a broad interval of $\sqrt{s}$ from
$19.4$ $GeV$ \cite{Sch81} up to $541$ $GeV$ \cite{CA93}. To
minimize the functional (\ref{eq1}), we used the least-square
method described in \cite{CERN}. The realistic result was obtained
when $\chi ^2/ndf\cong 1$ was achieved. Before using the \textit{a
priori} choice of $m_k$ in accordance with (\ref{eq7}), we set
$m_k=1$, assuming that nothing is known about $\frac{d\sigma
}{dt}$. In this case the use of the maximum entropy method just
reproduces the experimental behaviour of $\frac{d\sigma }{dt}$
with slight changes of the slope. The results of the calculations
with the \textit{a priori} distribution described above are shown
in Figures 1-3 for the ratios

\begin{figure}
\begin{center}
\includegraphics[scale=0.5]{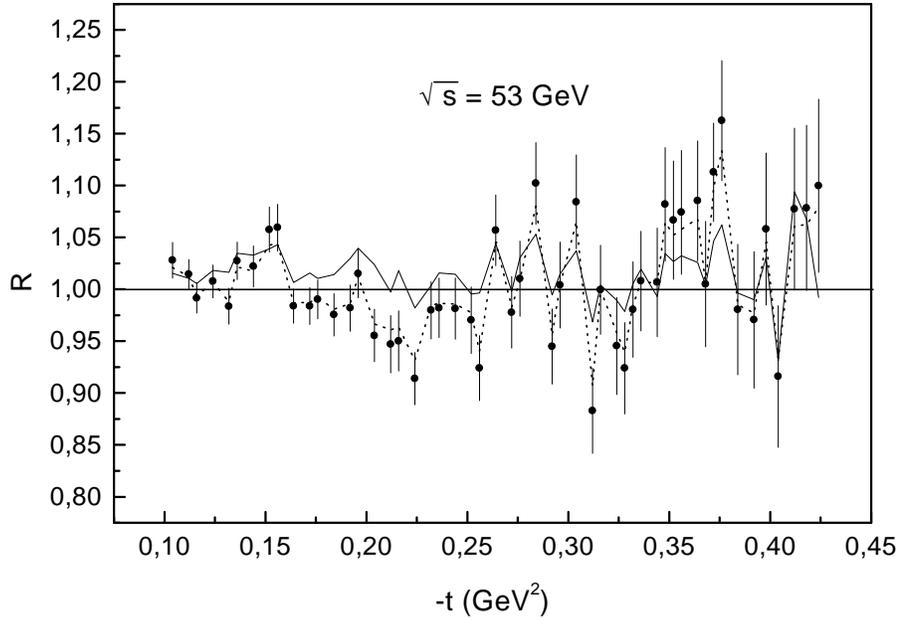}
\caption[]{Same as Figure 1 but for the data at $53$\ $GeV$\
\cite{BB72}. Here $\chi ^2/ndf\cong 1.01$ and $\lambda =0.23\cdot
10^6$ (solid curve). The dashed curve corresponds to the fit with
artificial errors (three times smaller), $\chi ^2/ndf\cong 0.98$
and $\lambda =0.35\cdot 10^6$.} \label{fig2}
\end{center}
\end{figure}

\begin{equation}
R_k^{entr}\left( t\right) =\frac{\left( \frac{d\sigma }{dt}\right)
_k^{entr}}{\left( \frac{d\sigma }{dt}\right) _k^{theor}},
\label{eq9}
\end{equation}

\begin{equation}
R_k^{exper}\left( t\right) =\frac{\left( \frac{d\sigma }{dt}\right)
_k^{exper}}{\left( \frac{d\sigma }{dt}\right) _k^{theor}}.  \label{eq10}
\end{equation}
and the relevant errors. Note that in all cases considered (at
$19.4$, $53$, and $541$ $GeV$) $\left( \frac{d\sigma }{dt}\right)
^{entr}$ exhibit those non-regularities which are present in
$\left( \frac{d\sigma }{dt}\right) ^{exper}$ albeit with smaller
amplitudes. At $53$ $GeV$ in spite of the decrease of the
amplitude of oscillations, the structure is still visible,
especially the local fluctuation at $t\sim -0.1$ $\left(
GeV/c\right) ^2$ discussed in \cite{AB76}. In contrast to the
first two cases, the data on the differential cross-section at
$541$ $GeV$ were taken from the region of interference of the
Coulomb and nuclear amplitudes used to determine the ratio of the
real part of the scattering amplitude to the imaginary part of the
scattering amplitude $\rho \left( s,t\right) $. The maximum
entropy method considerably reduces the amplitude of oscillations
in $\rho \left( s,t\right) $. By means of Fourier analysis we find
some characteristic periods which generally confirm the
calculation of \cite{PG97,KFT97}. However, the value of $\rho
\left( s,t\right)$ calculated on the basis of data processed by
using the maximum entropy method leads to such a drastic decrease
of the oscillation amplitude in $\rho \left( s,t\right)$ (see
Figure 4), that this gives one almost no chance to obtain the
value of $\rho \left( s,t=0\right)$ which noticeably differs from
the experimental value of $0.135$ (compare Figure 3 in \cite{PG97}
and Figure 4 of the present paper).

\begin{figure}
\begin{center}
\includegraphics[scale=0.5]{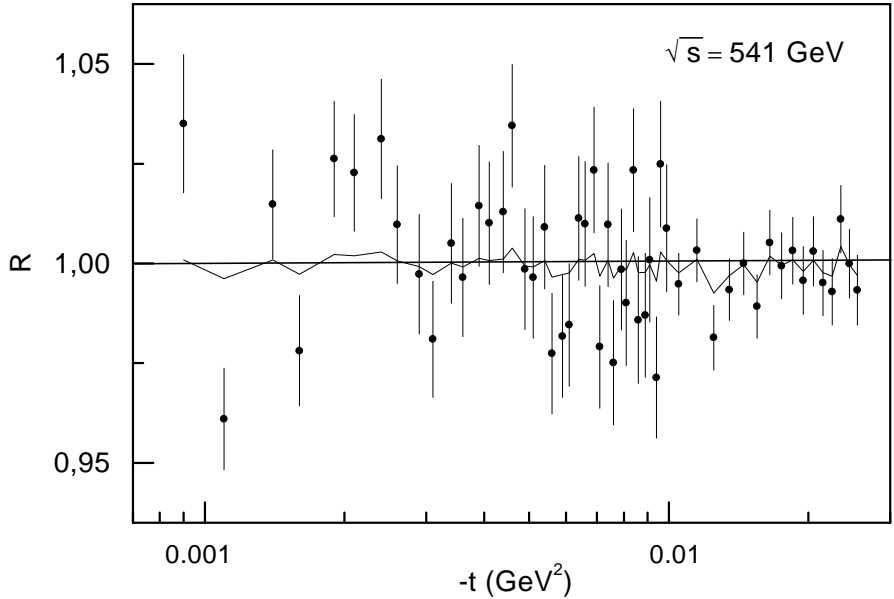}
\caption[]{Same as Figure 1 but for the UA4/2 data \cite{CA93}.
Here $\chi ^2/ndf\cong 1.04$ and $\lambda =0.35\cdot 10^6$.}
\label{fig3}
\end{center}
\end{figure}

Finally we made a numerical experiment by artificially reducing
three times the experimental errors of the data of \cite{BB72}. In
this case the amplitude of oscillations increased almost by an
order of magnitude (Figure 1) which stresses the importance of
efforts towards decreasing experimental errors in real
measurements.

\section{Concluding remarks}

We conclude that in all cases analysed the existing data reveal oscillations
satisfying quite realistic reliability criteria. Further studies for the
small-period structure in $t$ at RHIC \cite{Gur96} and LHC (see TOTEM
project \cite{MB96}) are highly desirable. To this end good statistics,
including measurements within the entire diffraction cone with a large
number points $(\sim 100)$ and small errors are necessary.

\medskip\

The authors are grateful to L. L. Jenkovszky for his constant interest in
this work. Fruitful discussions with A. V. Snegursky are gratefully
acknowledged.

\begin{figure}
\begin{center}
\includegraphics[scale=0.5]{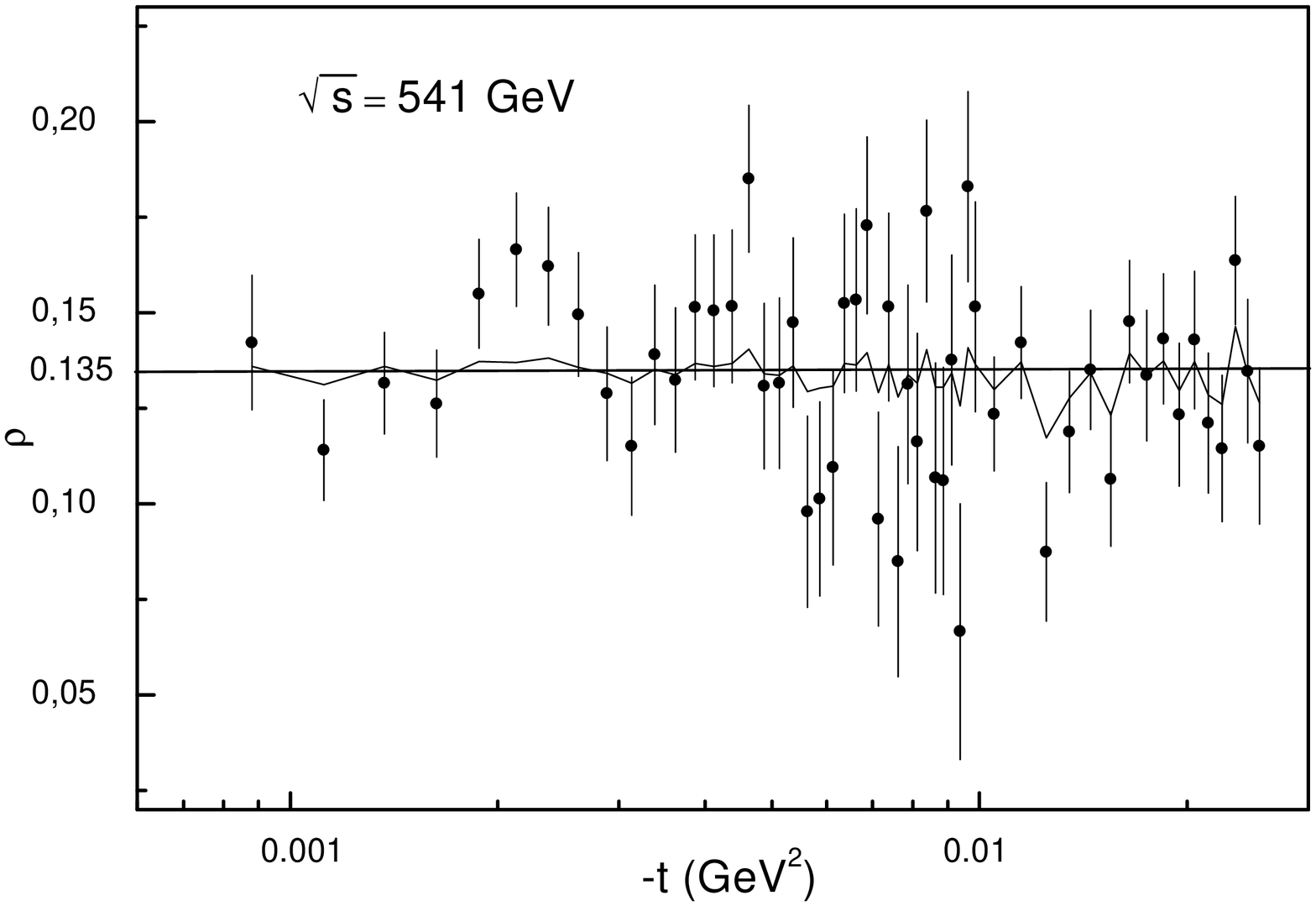}
\caption[]{The ratio $\rho \left( s,t\right)$ as a function of
$\sqrt{t}$ at $541$ $GeV$ (bars) calculated directly from the
experimental data with help

$\rho (s,t)=\frac 1{ImF_N}\left( \left[ 16\pi \frac{d\sigma
}{dt}-\left( ImF_C+ImF_N\right) ^2\right] ^{1/2}-ReF_C\right)$.

The solid line shows the calculations with the differential
cross-section corresponding to the solid line in Figure 3.}
\label{fig4}
\end{center}
\end{figure}

\newpage\

\end{document}